\documentclass[12pt]{article} 
\usepackage{graphicx}
\graphicspath{{Fig/}}
\usepackage{amsmath}
\numberwithin{equation}{section}
\usepackage{amssymb}
\usepackage{dsfont}
\usepackage{hyperref}
\usepackage[toc,page]{appendix}
\usepackage{caption}
\usepackage{titlesec}
\titleformat{\section}
  {\normalfont\fontsize{15}{18}\bfseries}{\thesection}{1em}{}
\DeclareMathOperator\sign{sgn}
\newcommand{\poiss}[1]{\left\{#1\right\}}
\newcommand{\lp}[0]{\nonumber \\}
\newcommand{\hil}[0]{\mathcal{H}}
\newcommand{\difm}[0]{Diff$(M)$}
\newcommand{\diffm}[0]{Diff$_{\mathcal{F}}(M)$}

\usepackage[export]{adjustbox}
\begin{document}

${}$\\
\begin{center}
\vspace{36pt}
{\Large \bf The initial-value formulation \\ \vspace{12pt} 
of the $\lambda$-R model}
\vspace{48pt}

{\sl L. Pires}\footnote{emails: l.pires@science.ru.nl, and luispenapires@gmail.com}

\vspace{24pt}

Radboud University, \\
Institute for Mathematics, Astrophysics and Particle Physics, \\
Heyendaalseweg 135, NL-6525 AJ Nijmegen, The Netherlands.\\ \vspace{2pt}

\vspace{72pt}

\end{center}

\begin{center}
{\bf Abstract}
\end{center}
We apply the conformal method to solve the initial value formulation of general relativity to the $\lambda$-R model, a minimal, anisotropic modification of general relativity with a preferred foliation and two local degrees of freedom. We obtain a generalised Lichnerowicz-York equation for the conformal factor of the metric and derive its properties. We show that the behaviour of the equation depends on the value of the coupling $\lambda$. In the absence of a cosmological constant, we recover the existence and uniqueness properties of the original equation when $\lambda>1/3$ and the trace of the momentum of the metric, $\pi$, is non-vanishing. For $\pi=0$, we recover the original Lichnerowicz equation regardless of the value of $\lambda$ and must therefore restrict the metric to the positive Yamabe class. The same restriction holds for $\lambda<1/3$, a case in which we show that the spatial Ricci scalar must also be large enough to guarantee the existence of at least one solution. Taking the equations of motion into account, this allows us to prove that there is in general no way of matching both constraint solving data and time evolution of phase space variables between the $\lambda$-R model and general relativity, thereby proving the the non-equivalence between the theories outside of the previously known cases $\lambda=1$ and $\pi=0$.

\newpage

\section{Introduction}
Despite the experimental success of general relativity \cite{GRExp}, there are still open questions when one uses it to describe the universe. Among them is the apparent necessity of introducing dark matter and dark energy to account for cosmological data, as well as the lack of a theory of quantum gravity. With regards to the latter, a common criterion to exclude a proposal for a theory of quantum gravity is whether or not its low energy physics are compatible, within current measuring accuracy, with observational data. This means that an acceptable low energy description should be close to general relativity when it concerns predictions from the latter that have been successfully tested. However, it can in principle not be equivalent to general relativity in its mathematical formulation. These constitute a good reason to consider modified theories of gravity at the purely classical level. The so-called $f(R)$-theories (see \cite{f(R)} for a review) are a well known example of such modified theories of gravity. In this text, we treat the initial value formulation of a different modification of Einstein's gravity, namely, the $\lambda$-R model \cite{BR1,DonJac,LP1,LP2}. It consists of a one-parameter family of gravitational theories with a preferred foliation ${\cal F}$ by leaves $\Sigma_t$ of constant time $t$, parametrised by $\lambda$ and including general relativity for the special value $\lambda=1$. It can be considered as a sand-alone modified theory of gravity or as a low energy, classical limit of the so-called Ho\v{r}ava-Lifshitz gravity, a proposal for a theory of quantum gravity introduced in \cite{HLG1} (see \cite{TS4,HLGRev} for two complementary reviews).

The fact that the theory possesses a preferred foliation as part of its background structure is in contrast with the general relativistic situation. In general relativity, one can write the four-dimensional spacetime manifold ${\cal M}$ as a product 
	\begin{equation}\label{1} {\cal M}=\mathds{R}\times\Sigma_t,\end{equation}
where $\Sigma_t$ is a smooth spacelike hypersurface and $\mathds{R}$ denotes a time direction\footnote{In what follows, we will have coordinates $x^i$ on the hypersurfaces $\Sigma_t$ and use $t\in\mathds{R}$ to denote the time coordinate.}. This follows from requiring global hyperbolicity of the spacetime ${\cal M}$. However, the decomposition \eqref{1} is not unique, as spacetime diffeomorphisms lead to physically equivalent solutions but do not necessarily preserve the foliation. In Ho\v{r}ava-Lifshitz gravity, and therefore also in the $\lambda$-R model, the notion of preferred foliation is implemented by restricting the symmetry group to be the subgroup of foliation-preserving diffeomorphisms \diffm. In the present text, we will follow the same approach we did in \cite{LP2} and not concern ourselves with the general viability of Ho\v{r}ava-Lifshitz gravity as a theory for quantum gravity, treating the $\lambda$-R model as a stand-alone classical theory of gravity instead.

The $\lambda$-R model is a constrained Hamiltonian system \cite{Dir,Sund,HT}, as is general relativity (see \cite{DG} for a recent review of the Hamiltonian treatment of general relativity). Thus, the closure of its constraint algebra \cite{BR1,LP1} implies that satisfying the constraints at an initial hypersurface $\Sigma_{t_0}$, with $t_0\in\mathds{R}$, guarantees that the time evolution of the fields is such that the constraints are satisfied at all times. The initial value formulation of the theory deals with the constraints at an initial hypersurface and aims to establish under what conditions can these constraints be satisfied, as well as which field components are constrained and which are freely specifiable. However, it does not concern itself with the time evolution equations. Because we want to be able to compare the four-dimensional spacetimes that are solutions to the $\lambda$-R model with those that are solutions of general relativity, we will nevertheless also take the time evolution equations of both theories into account.

To solve the initial value formulation of the $\lambda$-R model, we will employ the conformal method developed to tackle the same problem in general relativity \cite{YM1,YM2}. The reason for this choice is that in general relativity the conformal method includes a coordinate choice which is necessarily present in the $\lambda$-R model as a tertiary constraint\footnote{This is only true for $\lambda\neq 1/3$, as the same condition is a primary constraint of the model in that case.}. More concretely, in the conformal method, one chooses the initial extrinsic curvature tensor $K_{ij}$ to consist of a transverse-traceless piece, denoted $K_{ij}^{TT}$, and a constant trace $K=g^{ij}K_{ij}$, which obeys the so-called constant mean curvature condition,
	\begin{equation} K\equiv K(t).  \label{CMC}\end{equation}
As was showed in \cite{BR1,DonJac,LP1}, in the $\lambda$-R model, eq.\ \eqref{CMC} must be imposed as a constraint. This means that the conformal method is particularly suited to compare both theories, even if the interpretation of eq.\ \eqref{CMC} is different in the two cases. The difference is that in general relativity, the condition is used to gauge fix a representative of an equivalent class under four-dimensional diffeomorphisms \difm, while in the $\lambda$-R model all solutions of the model must satisfy eq.\ \eqref{CMC}.

This article is organised as follows. In the remainder of this section, we will recall the main features of the $\lambda$-R model, as well as the results pertaining its constraint algebra. Next, in Sec.\ \ref{sec:gr}, we review the conformal method applied to general relativity. This discussion is split in two parts, one focused on the Lichnerowicz equation, developed for $K=0$, and the other focused on the Lilchnerowicz-York equation, obtained when $K\neq 0$. Besides the historical relevance of the former discussion, the results for $K=0$ in general relativity turn out to be useful for the discussion of the conformal method in the context of the $\lambda$-R model. That analysis is then performed in Sec.\ \ref{IVLR}, where we obtain a modified Lichnerowicz-York equation, and subsequently determine the conditions under which it has solutions and whether those solutions are unique. The section is finalised with subsection \ref{Lapse-Fix}, in which we discuss the properties of the lapse-fixing equation, which in the $\lambda$-R model is a constraint and therefore must be part of the initial value formulation of the model. Finally, in Sec.\ \ref{Disc} we analyse the results of Sec.\ \ref{IVLR} and use them to compare the model to general relativity.

\subsection{The $\lambda$-R model}\label{ssec:lR}
In what follows, we use the $3+1$ ADM\footnote{The formulation is named after its authors: Richard Arnowitt, Stanley Deser and Charles Misner.} decomposition of the metric \cite{adm}, with line element
	\begin{equation}\label{line}
	ds^2=-N^2dt^2+g_{ij}\left(dx^i+N^idt\right)\left(dx^j+N^jdt\right),
	\end{equation}
where $g_{ij}(x,t)$ is the spatial metric on $\Sigma_t$, $N(x,t)$ is the lapse function, and $N^i(x,t)$ the shift three-vector. In general relativity, this formulation is particularly suited to treat the theory as a dynamical system, and it can be shown that the equations of motion for the four-dimensional metric $^{(4)}g_{\mu\nu}$ split into constraints\footnote{The equations of motion for the lapse $N$ and the shift $N^i$.}, and evolution equations, the equations of motion for the spatial metric $g_{ij}$ \cite{num}. In this formulation, the Einstein-Hilbert action is given by
	\begin{equation}\label{SEH}
	S_{EH}=\frac{1}{16\pi G_N}\int dt\int d^3x\,\sqrt{g}\,N\left(K^{ij}K_{ij}-K^2+{\cal R}-2\Lambda\right),
	\end{equation}
where we have discarded any boundary terms, $G_N$ denotes Newton's constant, $\Lambda$ the cosmological constant, ${\cal R}$ the Ricci scalar on $\Sigma_t$, and $K_{ij}$ the extrinsic curvature tensor,
	\begin{equation}
	K_{ij}=\frac{1}{2N}\left(\dot{g}_{ij}-\nabla_iN_j-\nabla_jN_i\right),
	\end{equation}
where the dot denotes a derivative with respect to time, and $\nabla_i$ denotes the spatial covariant derivative in the $i$-direction with respect to the spatial metric $g_{ij}$. It is possible to write the kinetic term of eq.\ \eqref{SEH} in terms of the Wheeler-DeWitt metric $G^{ijkl}$,
	\begin{equation}\label{OGwdw}
	G^{ijkl}=\frac{1}{2}\left(g^{ik}g^{jl}+g^{il}g^{jk}\right)-g^{ij}g^{kl},
	\end{equation}
which is an ultralocal\footnote{A metric $G^{ijkl}$ is said to be ultralocal if it does not depend on spatial derivatives of the (inverse) metric $g^{ij}$.} metric in the space of three-metrics on a hypersurface $\Sigma_t$, ${\rm Riem}\,\Sigma_t$. To distinguish $G^{ijkl}$ from the individual elements on ${\rm Riem}\,\Sigma_t$, $G^{ijkl}$ is usually referred to as a supermetric. Using this supermetric, the Einstein-Hilbert action becomes
	\begin{equation}\label{SEH2}
	S_{EH}=\frac{1}{16\pi G_N}\int dt\int d^3x\,\sqrt{g}\,N\left(K_{ij}G^{ijkl}K_{kl}+{\cal R}-2\Lambda\right).
	\end{equation}
The $\lambda$-R model can be obtained by substituting the Wheeler-DeWitt metric in eq.\ \eqref{SEH2} by a generalised supermetric $G^{ijkl}_{\lambda}$,
	\begin{equation}\label{Gwdw}
	G^{ijkl}_{\lambda}=\frac{1}{2}\left(g^{ik}g^{jl}+g^{il}g^{jk}\right)-\lambda g^{ij}g^{kl},
	\end{equation}
which constitutes the most general ultralocal supermetric on ${\rm Riem}\,\Sigma_t$ and was first introduced in \cite{BdW}. The action of the $\lambda$-R model is therefore given by
	\begin{equation}
	S_{\lambda}=\frac{1}{16\pi G_N}\int dt\int d^3x\,\sqrt{g}\,N\left(K_{ij}G^{ijkl}_{\lambda}K_{kl}+{\cal R}-2\Lambda\right).
	\end{equation}
Note that we are adhering to the line element \eqref{line} and subsequent definition of fields in the definition of the $\lambda$-R model, which implies that the lapse is also regarded as a function of spacetime. The name of the $\lambda$-R model was coined in \cite{BR1}, where the model as we define it was first studied in the context of Ho\v{r}ava-Lifshitz gravity. The reason for the ``R'' in its name was to distinguish from other models of Ho\v{r}ava-Lifshitz gravity that have other terms besides the Ricci scalar in its potential. While we do not denote the Ricci scalar by $R$ as usual, we will refer to the model by its common name.

In both the revision that follows and in Sec.\ \ref{IVLR}, where we apply the conformal method to the $\lambda$-R model, we will work in the Hamiltonian formalism. Since the equations defining the Legendre transformation will be important in Sec.\ \ref{IVLR}, let us state them here 
	\begin{subequations}\label{b12}\begin{align}
	&\pi^{ij}:=\frac{\delta S}{\delta\dot{g}_{ij}}=\sqrt{g}\,G^{ijkl}_{\lambda}K_{kl},
	\label{b12a}\\
	&\phi:=\frac{\delta S}{\delta\dot{N}}=0,\qquad \phi_i:=\frac{\delta S}{\delta\dot{N}^i}=0.
	\label{b12b}
	\end{align}\end{subequations}
Note that it is only possible to invert eq.\ \eqref{b12a} for $\pi^{ij}$ if $\lambda\neq 1/3$, because the inverse generalised Wheeler-DeWitt metric only exists under that condition,
	\begin{equation}
	G_{ijkl}^{\lambda}=\frac{1}{2}\left(g_{ik}g_{jl}+g_{il}g_{jk}\right)-\frac{\lambda}{3\lambda-1}g_{ij}g_{kl}.
	\end{equation}
When $\lambda=1/3$, the $\lambda$-R model becomes Weyl invariant and $\pi=0$ becomes a primary constraint. In what follows, we will not concern ourselves with this case. For the range of values of $\lambda$ that we are considering, the primary constraints of the $\lambda$-R model are the same as those of general relativity, i.e., the vanishing of $\phi$ and $\phi_i$, the momenta of the lapse $N$ and the shift $N^i$ respectively. Ignoring any boundary contributions, the associated total Hamiltonian is 
	\begin{equation}\label{b16}
	H_{tot}=\int d^3x\,\left(N\hil_{\lambda}+N^i\hil_i+\alpha\phi+\alpha^i\phi_i\right),
	\end{equation}
where $\alpha$ and $\alpha^i$ are Lagrange multipliers respectively associated with $\phi$ and $\phi_i$, while $\hil_{\lambda}$ and $\hil_i$ are the following functionals of the metric $g_{ij}$ and its momentum $\pi^{ij}$,
	\begin{subequations}\label{b15}\begin{align}
	\label{b15a}\hil_{\lambda}&=
	\frac{1}{\sqrt{g}}\,G^{\lambda}_{ijkl}\,\pi^{ij}\pi^{kl}-\sqrt{g}\,\left({\cal R}-2\Lambda\right),\\
	\label{b15b} \hil_i&=-2g_{ij}\,\nabla_k\,\pi^{jk}.
	\end{align}\end{subequations}
Because the total Hamiltonian is linear in the lapse and in the shift, it follows that the secondary constraints are a $\lambda$-dependent Hamiltonian constraint,
	\begin{equation}
	\dot{\phi}\approx 0\;\Rightarrow \poiss{\phi,H_{tot}}=\hil_{\lambda}\approx 0,
	\end{equation}
and the usual momentum constraints of general relativity,
	\begin{equation}
	\dot{\phi}_i\approx 0\, \Rightarrow \poiss{\phi_i,H_{tot}}=\hil_i\approx 0.
	\end{equation}
Next in the Dirac algorithm, one demands that the secondary constraints $\hil_{\lambda}\approx 0$ and $\hil_i\approx 0$ are preserved in time. The Poisson bracket between the momentum constraints and the total Hamiltonian vanishes weakly, as it does in general relativity, but the same does not hold for the Hamiltonian constraint. As was shown in \cite{BR1,DonJac}, the demand that $\hil_{\lambda}$ is preserved in time requires the presence of a tertiary constraint to be met, namely,
	\begin{equation}\label{CMC-PS}
	\omega:=\pi-a(t)\sqrt{g}\approx 0,
	\end{equation}
which is the phase-space version of the constant mean curvature condition \eqref{CMC}. We further showed in \cite{LP1} that preserving eq.\ \eqref{CMC-PS} in time introduces a lapse-fixing equation as a quaternary constraint\footnote{In \cite{LP1}, we imposed $\dot{a}=0$ for simplicity. We are not performing the same simplification in the current text.},
	\begin{equation}\label{tert}
	{\cal M}:=D_{\lambda}N-\frac{\dot{a}}{2}\sqrt{g}\approx 0,
	\end{equation}
where $D_{\lambda}$ is the following differential operator
	\begin{equation}
	D_{\lambda}:=\sqrt{g}\left({\cal R}-3\Lambda+\frac{a^2}{2\left(3\lambda-1\right)}-\nabla^2\right).
	\end{equation}
The Dirac algorithm ends when one demands that ${\cal M}$ is preserved in time, thus obtaining an equation for the Lagrange multiplier $\alpha$.

In Sec.\ \ref{IVLR}, we will not only compare the solutions to the constraints of general relativity and the $\lambda$-R model, but also the way they evolve in time. We will therefore need the equations of motion for the metric $g_{ij}$ and its momentum $\pi^{ij}$, 
	\begin{subequations}\label{eom}\begin{align}
	\dot{g}_{ij}=&\,\frac{2N}{\sqrt{g}}\left(\pi_{ij}-\frac{\lambda}{3\lambda-1}\pi\, g_{ij}\right)+g_{ik}\nabla_j\,N^k+g_{jk}\nabla_i\,N^k\,,\label{145a}\\
	\dot{\pi}^{ij}=&-\frac{2N}{\sqrt{g}}\left(g_{kl}\pi^{ik}\pi^{jl}-\frac{\lambda}{3\lambda-1}\pi\pi^{ij}\right)-N\sqrt{g}\left({\cal R}^{ij}-g^{ij}\left(\Lambda-\frac{\lambda}{2\left(3\lambda-1\right)}\,a^2\right)\right)\lp
	&+\sqrt{g}\,g^{ik}g^{jl}\nabla_k\nabla_l N+\nabla_a\left(N^a\,\pi^{ij}\right)-\pi^{ai}\,\nabla_a\,	N^j-\pi^{aj}\,\nabla_a\,N^i\,.\label{145b}
	\end{align}\end{subequations}
Note that for $\pi\neq 0$ to be an acceptable solution to the tertiary constraint $\omega\approx 0$, one must consider hypersurfaces that are either asymptotically null or compact, without boundary \cite{LP1}. As can be seen in eqs.\ \eqref{b15a},\ \eqref{tert}, and\ \eqref{eom}, when $\pi=0$, $\lambda$ drops out of both the constraints and the equations of motion. Nevertheless, if we consider not only asymptotically flat hypersurfaces, but also asyptotically null or compact and without boundary ones, the $\lambda$-R model does not reproduce a gauge fixed version of general relativity. 

In reference \cite{LP2}, we obtained non-static spherically symmetric solutions of the model. These agree with the Schwarzschild solution for asymptotically flat boundary conditions, but predict a non-vanishing four-dimensional curvature, which depends not only on $\lambda$, but also on quantities that are gauge-parameters of general relativity. This concrete example demonstrates that there are circumstances in which the $\lambda$-R model does not agree with general relativity and the spacetime geometry becomes $\lambda$-dependent. However, the equations of motion \eqref{eom} carry gauge-redundancies which makes it impossible to use them to precise how $\lambda$ influences the solutions, while the solutions presented in \cite{LP2} were obtained in a reduced setting. As we will show in Sec.\ \ref{IVLR}, the initial value formulation of the model allows one to disentangle the influence of $\lambda$ in the geometry and to make precise and general comparisons with general relativity.

\section{The conformal method in general relativity}\label{sec:gr}
In this section, we briefly review the the conformal method in general relativity (see \cite{Rd} for a recent, more comprehensive discussion). This discussion is divided in two parts. In subsection \ref{Max}, we present the conformal method for maximal slicing coordinates, which yields Lichnerowicz equation, while in subsection \ref{sec:YLe} we present the generalisation to constant mean curvature coordinates and the associated Lichnerowicz-York equation. This review is presented in the Lagrangian setting, as that is how the original results were derived. For the same reason, we will not consider a cosmological constant in this section. This is in contrast with how we will apply the method to the $\lambda$-R model in Sec.\ \ref{IVLR}, where we will work in the Hamiltonian formalism and consider a non-vanishing cosmological constant.

Recall the form of the Hamiltonian and momentum constraints of general relativity,
	\begin{subequations}\label{GRcons}\begin{align}
	&\hil\left[g_{ij},K_{ij}\right]:={\cal R}-K_{ij}K^{ij}+K^2=0,\label{c1a}\\
	&\hil_i\left[g_{ij},K_{ij}\right]:=\nabla_i\left(K^{ij}-g^{ij}K\right)=0.\label{c1b},
	\end{align}\end{subequations}
where we have once again introduced the shorthands $\hil$ and $\hil_i$ to respectively denote the Hamiltonian and momentum constraints. Because these constraints must be satisfied at the initial hypersurface $\Sigma_{t_0}$, the initial values of the metric $g_{ij}(t_0,x^i)$ and the extrinsic curvature tensor $K_{ij}(t_0,x^i)$, cannot be freely specified. They have to be such that eqs.\ \eqref{GRcons} are satisfied. 

In what fallows, we will deal with two sets of variables related by a conformal transformation with conformal factor $\phi$. One is the so-called set of initial data, which we will denote by the usual symbols for the metric and the extrinsic curvature, $\left(g_{ij},K_{ij}\right)$. The second set consists of the so-called constraint-solving data and we will denote its elements by barred versions of the variables, $\left(\bar{g}_{ij},\bar{K}_{ij}\right)$. The same distinction between sets of data will be used in Sec.\ \ref{IVLR} with the extrinsic curvature replaced by the momentum density tensor $\pi^{ij}$.

The method can be described as follows. One first chooses an extrinsic curvature tensor $K_{ij}$ which satisfies $\hil_i\left[g_{ij},K_{ij}\right]= 0$ and such that, after a specific conformal transformation, the transformed constraints $\hil_i\left[\bar{g}_{ij},\bar{K}_{ij}\right]=0$ are also satisfied. One then writes the Hamiltonian constraint in terms of the constraint-solving data $\hil\left[\bar{g}_{ij}\bar{K}_{ij}\right]= 0$, and substitutes the expression of these variables in terms of initial data and the conformal factor. This yields $\hil\left[\phi,g_{ij},K_{ij}\right]= 0$, which can be shown to be an equation for the conformal factor\footnote{There are other interpretations to this equation, which require a reformulation of what constitutes initial data. These are discussed in \cite{Rd}.} $\phi$. When, for given initial data, the equation for $\phi$ has a solution, it is the constraint-solving data that solves both the momentum and the Hamiltonian constraints, hence the name. Note that for given constraint-solving data, there is a family of sets of initial data related to it by conformal transformations.

\subsection{Maximal slicing and the Lichnerowicz equation}\label{Max}
The first step in the development of the model follows the work of French mathematician Lichnerowicz \cite{AL}, where it was noticed that when the extrinsic curvature tensor $K_{ij}$ is traceless an transverse with respect to the metric $g_{ij}$, that is,
	\begin{equation} \label{TTdef}
	K=g^{ij}K_{ij}=0\, ,\qquad\mbox{and}\qquad \nabla_i K^{ij}=0\, ,
	\end{equation}
the momentum constraints $\hil_i\approx 0$ \eqref{c1b} are solved. As a consequence of solving those constraints by choosing an extrinsic curvature satisfying eq.\ \eqref{TTdef}, they decouple from the Hamiltonian constraint. From now on, we will denote any tensor (or tensor density) $A_{ij}$ satisfying eq.\ \eqref{TTdef} as $A_{ij}^{TT}$, where $TT$ stands for transverse-traceless.

We now consider an extrinsic curvature tensor $K_{ij}$ which satisfies eq.\ \eqref{TTdef} and denote it by $K_{ij}^{TT}$ . It is possible to define the conformal transformation of both the metric and the transverse-traceless extrinsic curvature to be such that the transformed extrinsic curvature $\bar{K}_{ij}$ is transverse-traceless with respect to the transformed metric $\bar{g}_{ij}$. It follows that the transformed tensors still solve the momentum constraints.

To understand what the choice of a transverse-traceless extrinsic curvature tensor entails, consider the following general but not-unique decomposition of $K_{ij}$,
	\begin{equation}\label{c2}
	K_{ij}=K_{ij}^{TT}+\nabla_iv_j+\nabla_jv_i-\frac{2}{3}g_{ij}\nabla_kv^k+\frac{1}{3}g_{ij}K,
	\end{equation}
where the vector $v_i$ describe the longitudinal components of the extrinsic curvature $K_{ij}$. Therefore, choosing $K_{ij}$ to be transverse-traceless with respect to the metric $g_{ij}$ implies setting both its longitudinal components and trace to zero. Using this choice of extrinsic curvature, the Hamiltonian constraint \eqref{c1a} reduces to
	\begin{equation}\label{c5}
	\hil\left[g_{ij},K_{ij}^{TT}\right]=0\Leftrightarrow {\cal R}-K_{ij}^{TT}K^{ij}_{TT}=0.
	\end{equation}
To proceed, we specify the form of the conformal transformation which preserves this choice of extrinsic curvature. For the metric $g_{ij}$, we define
	\begin{equation}\label{c6}
	\bar{g}_{ij}=\phi^4g_{ij},
	\end{equation}
where $\phi$ is a function on $\Sigma_{t_0}$, which we define to be everywhere strictly positive. We want to ensure that this transformation is such it maps the transverse-traceless extrinsic curvature to a tensor which is transverse-traceless with respect to $\bar{g}_{ij}$ \eqref{c6}, that is
	\begin{equation}
	\left(g_{ij},K_{ij}^{TT}\right)\underset{\phi}{\longmapsto}\left(\bar{g}_{ij},\bar{K}_{ij}^{TT}\right),
	\end{equation}
which is achieved if we define the transformation of the transverse-traceless extrinsic curvature as
	\begin{equation}\label{c7}
	\bar{K}_{ij}^{TT}=\phi^{-2}K_{ij}^{TT}.
	\end{equation}
The reason for the choice of power of $\phi$ in eq.\ \eqref{c7} is that, in three dimensions, it can be shown that defining
	\begin{equation}\bar{K}_{ij}^{TT}=\phi^{n}K_{ij}^{TT},
	\end{equation}
then the left-hand side is transverse-traceless with respect to $\bar{g}_{ij}$ \eqref{c6} if and only if $n=-2$. Under eq.\ \eqref{c6}, the Ricci scalar curvature ${\cal R}$ transforms as
	\begin{equation}\label{c8}
	\bar{{\cal R}}=\phi^{-4}{\cal R}-8\,\phi^{-5}\nabla^2\phi.
	\end{equation}
Note that the only derivative operator entering the right-hand side is the Laplacian. This is the motivation for the definition of eq. \eqref{c6} with that specific power of $\phi$ in the first place, as it can be shown that it is the only choice for which this property holds.

In the next step, we write the Hamiltonian constraint eq.\ \eqref{c5} for the barred variables (constraint-solving data) and substitute these variables by the initial data and conformal factor according to eqs. \eqref{c6}, \eqref{c7}, and \eqref{c8},
	\begin{equation}\label{c9}
	\hil\left[\bar{g}_{ij},\bar{K}_{ij}^{TT}\right]=\hil\left[\phi,g_{ij},K_{ij}^{TT}=0\right]\Leftrightarrow 8\,\nabla^2\phi={\cal R}\,\phi-\phi^{-7}K^{ij}_{TT}\,K^{TT}_{ij},
	\end{equation}
which is the Lichnerowicz equation.

To address the solutions of the equation, it is preferable to work with a constant Ricci scalar ${\cal R}$. To see how this is in general possible, we refer to appendix \ref{app yam} at the end of this article, where we discuss the Yamabe classification. Note that this classification can only be applied in full generality to compact hypersurfaces. Hence, the results that follow in the remainder of this article, apply to compact hypersurfaces and to a subset of all asymptotically null and flat hypersurfaces. From now on, we assume that the Ricci scalar associated with the set of initial data is a constant of a given sign. It is either positive, negative, or vanishing depending on whether $\left(\Sigma_{t_0},g_{ij}\right)$ belongs to the positive, negative, or vanishing Yamabe class, respectively.

Let us return to the Lichnerowicz equation \eqref{c9} and integrate it over $\Sigma_{t_0}$. Due to Stokes' theorem, the integral of the left-hand side vanishes,
	\begin{equation}\label{c12}
	8\int d^3x \sqrt{g}\, \nabla^2\phi =0.
	\end{equation}
This implies that the integral of the right-hand side must also vanish,
	\begin{equation}\label{c13}
	\int d^3x\sqrt{g}\, \phi {\cal R}= \int d^3x\sqrt{g}\,\phi^{-7}K_{TT}^{ij}K_{ij}^{TT}.
	\end{equation}
Note that $\phi>0$ by definition, $K_{ij}^{TT}K_{TT}^{ij}\geq 0$, and ${\cal R}$ is a constant. Hence, the equation only admits solutions when $g_{ij}$ belongs to the positive Yamabe class. Because the Yamabe classification is a conformal invariant, this means that when the extrinsic curvature is transverse-traceless, the Hamiltonian constraint can only be solved by metrics belonging to the positive Yamabe class.

\subsection{Constant mean curvature and the Lichnerowicz-York equation}
\label{sec:YLe}
The next big development of the conformal method is based on the work of York in \cite{JY1,JY2,JY3}. In these references, it was shown that the conformal invariance of the solution of the momentum constraints is kept intact if the extrinsic curvature tensor also includes a non-vanishing - albeit spatially constant  - trace. This alternative choice of extrinsic curvature tensor has the advantage that when the Hamiltonian constraint is transformed into an equation for the conformal factor, the so-called Lichnerowicz-York equation, there are solutions for an almost unrestricted choice of initial data (we explain what is meant by ``almost'' below). 

The constant trace condition is the so-called constant mean curvature condition, 
	\begin{equation}\label{cmcdef}
	\nabla_iK=\partial_iK=0,
	\end{equation}
and, instead of choosing extrinsic curvature to consist of a symmetric transverse-traceless tensor \eqref{TTdef}, we choose it to also include a non-vanishing trace $K$ that satisfies eq. \eqref{cmcdef},
	\begin{equation}\label{c14}
	K_{ij}=K_{ij}^{TT}+\frac{1}{3}g_{ij}K.
	\end{equation}
We then substitute this choice of extrinsic curvature $K_{ij}$ \eqref{c14} into the momentum constraints $\hil_i$ \eqref{c1b}. It is straightforward to see that the trace term cancels due to the constant mean curvature condition while the $K_{ij}^{TT}$-term cancels by virtue of the transverse-traceless property. Eq. \eqref{c14} is therefore a solution to the momentum constraints. In subsection \ref{Max} above, we defined the transformation of $K_{ij}^{TT}$ under $\phi$ to be such that $\bar{K}_{ij}^{TT}$ remains transverse-traceless with respect to the transformed metric $\bar{g}_{ij}$. For the same reason, we define the transformation of the trace $K$ to be such that $\bar{K}$ obeys the constant mean curvature, i.e., we define $\bar{K}=K$. With these choices, the constraint solving data is related to the initial data by
	\begin{equation}\label{c15}
	\bar{g}_{ij}=\phi^4 g_{ij},\qquad \bar{K}_{ij}^{TT}=\phi^{-2}K_{ij}^{TT},\qquad \bar{K}=K.
	\end{equation}
This implies that the original extrinsic curvature tensor $K_{ij}$ does not transform homogeneously under this map. We should therefore not consider the initial data as a pair of a symmetric tensors $K_{ij}$ and $g_{ij}$, but rather a trio consisting of the metric $g_{ij}$, a symmetric transverse-traceless tensor $K_{ij}^{TT}$, and a constant scalar $K$, which should all be specified independently.

As before, we write the Hamiltonian constraint \eqref{c1a} for the constraint solving data and substitute the conformal transformations \eqref{c15},
	\begin{align}
	&\hil\left[\bar{g}_{ij},\bar{K}_{ij}^{TT},\bar{K}\right]=\hil\left[\phi,g_{ij},K_{ij}^{TT},K\right]=0,\lp 
	\Leftrightarrow \,\,& 8\nabla^2\phi={\cal R} \phi-\phi^{-7}K^{ij}_{TT}K^{TT}_{ij}+\frac{2}{3}\phi^5 K^2. \label{c16}
	\end{align}
As we did in subsection \ref{Max} above, we assume that ${\cal R}$ is a constant . To discuss the existence and uniqueness of solutions of the Lichnerowicz-York equation \eqref{c16}, it is useful to think of its right-hand side as a polynomial in $\phi$,
	\begin{equation}\label{c17}
	P(\phi):={\cal R}\phi-\phi^{-7}{\cal A}+\frac{2}{3}\phi^5 K^2,
	\end{equation}
where ${\cal A}=K^{ij}_{TT}K^{TT}_{ij}\geq 0$. For the reason alluded to when addressing the Lichnerowicz equation, the integral of the left-hand side of eq. \eqref{c16} vanishes. As a consequence, so does the integral of $P(\phi$) over $\Sigma_{t_0}$,
	\begin{equation}\label{c18}
	\int d^3x\sqrt{g} \, P(\phi)=0.
	\end{equation}
This implies that the polynomial must have at least one zero. Provided ${\cal A}\neq 0$ everywhere on $\Sigma_{t_0}$, the asymptotic behaviour of $P(\phi)$ is compatible with this requirement, 
	\begin{equation}\label{c19}
	\lim_{\phi\rightarrow \,0^+}P(\phi)=-\infty,\qquad \lim_{\phi\rightarrow +\infty}P(\phi)=+\infty.
	\end{equation}
Unlike in the maximal slicing $K=0$ case, the Yamabe class of $\left(\Sigma_{t_0},g_{ij}\right)$ is not relevant to the existence of at least one zero. However, the existence of $\phi_0$ such that $P(\phi_0)=0$ is not enough to guarantee the existence of a solution. An exception occurs when ${\cal A}$ is a constant, in which case $\phi_0$ is in itself a solution to the Lichnerowicz-York equation. In reference \cite{YM2}, York and Murchadha prove two theorems that not only guarantee the existence of a solution, but also show that it is almost always unique. By almost always unique, it is understood that the cases for which the solution is not unique have zero measure in the set of all choices of initial data. The first theorem states that the Lichnerowicz-York equation has a positive, bounded solution $\phi_{sol}$ if there exist two positive constants $\phi_-<\phi_+$ such that,
	\begin{equation}\label{c20}
	\left.\begin{array}{c}
	P(\phi_-)<0 \\ P(\phi_+)>0
	\end{array}\right\}\forall x\in\Sigma_{t_0}.
	\end{equation}
The theorem further ensures that the solution $\phi_{sol}$ lies in the interval $\left(\phi_-,\phi_+\right)$.

It is also possible to show, as done in the same text \cite{YM2}, that $P(\phi)$ only has one zero. Since both ${\cal R}$ and $K$ are constants on $\Sigma_{t_0}$, eq. \eqref{c19} shows that the polynomial behaves as required by \eqref{c20} around the zero. This means that we are guaranteed a bounded, positive solution as long as ${\cal A}$ is bounded. This is a reasonable restriction on the initial data, since ${\cal A}\rightarrow\infty$ would describe an unphysical situation. The second theorem proved in \cite{YM2} guarantees that the solution $\phi_{sol}$ is unique except in the trivial case of ${\cal A}=K=0$ everywhere.

\section{The initial value formulation of the $\lambda$-R model}\label{IVLR}
As we reviewed in subsection \ref{ssec:lR}, the constraint algebra of the $\lambda$-R model closes for spatial hypersurfaces that are either compact and without boundary, asymptotically flat, and asymptotically null. It follows that if all the constraints are satisfied at some initial hypersurface $\Sigma_{t_0}$, then the time evolution equations \eqref{eom} guarantee that the constraints will be satisfied for all times.

To apply the conformal method reviewed in Sec.\ \ref{sec:gr} to the $\lambda$-R model, the first step consists of a choice of momentum tensor which solves the momentum constraints. It is therefore convenient to recall the functional form of the momentum and Hamiltonian constraints \eqref{b15}
	\begin{subequations}\label{cc4}\begin{align}
	&\hil_i=-2g\,_{ij}\nabla_k\,\pi^{jk}\approx 0\,,\label{cc4a}\\
	&\hil_{\lambda}=\frac{1}{\sqrt{g}}\,G^{\lambda}_{ijkl}\,\pi^{ij}\pi^{kl}
	-\sqrt{g}\left({\cal R}-2\Lambda\right)\approx 0\,.
	\end{align}\end{subequations}
Further recall that in the $\lambda$-R model, the tertiary constraint $\omega\approx 0$ \eqref{CMC-PS} is solved by the constant mean curvature condition
	\begin{equation}\label{cc2}
	 \pi=a(t)\sqrt{g}\,,
	\end{equation}
where $a(t)$ is a constant for each $\Sigma_t$. This is the phase space version of the constant-$K$ condition which is a necessary condition to obtain the Lichnerowicz-York equation \eqref{c16}. As we saw in Sec.\ \ref{sec:gr}, the first step in the conformal method consists of a choice of decomposition of the extrinsic curvature. Because we are working in the Hamiltonian formulation, the equivalent step is to choose a momentum tensor density $\pi^{ij}$. For our purposes, we choose a $\pi^{ij}$ whose only non-vanishing pieces are its trace $\pi$ and its transverse traceless components $\pi^{ij}_{TT}$,
	\begin{equation}\label{cc3}
	\pi^{ij}=\pi^{ij}_{\,TT}+\frac{1}{3}\,g^{ij}\pi\,,
	\end{equation}
where $\pi$ satisfies eq.\ \eqref{cc2}. 

Recall from eq.\ \eqref{b12a} that the definition of $\pi^{ij}$ through the Legendre transformation is $\lambda$-dependent. Because we want to study the way the solutions of the Hamiltonian constraint are sensitive to the value of $\lambda$, it is important to understand how the initial data $\pi^{ij}_{TT}$ and $\pi$ depends on $\lambda$. Taking the trace of eq.\ \eqref{b12a}, we obtain
	\begin{equation}\label{Pi-K}
	\pi=\sqrt{g}\left(1-3\lambda\right)K,
	\end{equation}
which is $\lambda$-dependent. Combining the decomposition \eqref{cc3} with eq.\ \eqref{Pi-K} and the Legendre transformation \eqref{b12a}, we obtain
	\begin{equation}
	\pi^{ij}_{TT}=\sqrt{g}K^{ij}_{TT},
	\end{equation}
which implies that $\pi^{ij}_{TT}$ is $\lambda$-independent and that the quantity ${\cal A}$, defined in subsection \ref{sec:YLe}, can be equivalently expressed in terms of the initial data of the $\lambda$-R model,
	\begin{equation}
	{\cal A}=K^{ij}_{TT}K_{ij}^{TT}=\frac{\pi^{ij}_{TT}\pi_{ij}^{TT}}{g},
	\end{equation}
where $g$ is the determinant of the spatial metric $g_{ij}$.

It also follows from the decomposition \eqref{cc3} with $\pi$ satisfying eq.\ \eqref{cc2} that this choice of $\pi^{ij}$ solves the momentum constraints $\hil_i\approx 0$. Substituting this decomposition into the Hamiltonian constraint $\hil_{\lambda}\approx 0$, it reads
	\begin{equation}\label{cc5}
	\hil_{\lambda}=\frac{1}{\sqrt{g}}\left(\pi^{ij}_{\,TT}\,\pi^{\,TT}_{ij}-\frac{1}{3}\frac{\pi^2}{3\lambda-1}\right)-\sqrt{g}\left({\cal R}-2\Lambda\right)\approx 0\,.
	\end{equation}
The momentum-space version of the extrinsic curvature transformations \eqref{c7} and $\bar{K}=K$ can be deduced by combining those relations with the conformal transformation of the metric \eqref{c6} and the Legendre transformation \eqref{b12a}, yielding
	\begin{subequations}\label{cc6}\begin{align}
	&\bar{\pi}^{ij}_{\,TT}=\phi^{-4}\pi^{ij}_{\,TT},\,\\
	&\bar{\pi}=\phi^6\pi\,.
	\end{align}\end{subequations}
Note that the transformation of $\pi$ reflects the fact that it is a density, i.e., it is not $\pi$ that transforms as a scalar but $\pi/\sqrt{g}$. 

Now that we have established the phase-space version of the conformal map, we write the Hamiltonian constraint as a functional of the constraint-solving data,
	\begin{equation}
	\hil_{\lambda}\left[\bar{g}_{ij},\bar{\pi}^{ij}_{TT},\bar{\pi}\right]\approx 0,
	\end{equation}
and substitute the barred variables by their expressions in terms of the conformal factor and initial data given in eqs.\ \eqref{c6} and \eqref{cc6},
	\begin{equation}\label{esta}
	\hil_{\lambda}\left[\phi,g_{ij},\pi^{ij}_{TT},\pi\right]\approx 0. 
	\end{equation}
After a few algebraic manipulations, eq.\ \eqref{esta} becomes the modified Lichnerowicz-York equation,
	\begin{equation}\label{cc7}
	8\nabla^2\phi=\phi {\cal R}-\,\phi^{-7}\,\frac{\pi^{ij}_{\,TT}\,\pi^{\,TT}_{ij}}{g}+\phi^5\left(\frac{1}
	{3\left(3\lambda-1\right)}\frac{\pi^2}{g}-2\Lambda\right).
	\end{equation}
Next, we study the existence of solutions of eq.\ \eqref{cc7}, treating separately the regimes in which the $\phi^5$-term on the right-hand side has different signs.
\subsection{The modified Lichnerowicz-York equation}
As happened in Sec.\ \ref{sec:gr}, if we integrate the left-hand side of the modified Lichnerowicz-York equation \eqref{cc7} over $\Sigma_{t_0}$, Stokes' theorem implies that the integral vanishes. It follows that the integral of the right-hand also vanishes. We thus write the right-hand side of the modified Lichnerowicz-York equation as a polynomial $P(\phi)$ in $\phi$,
	\begin{equation}\label{cc8}
	P(\phi):=\phi \,{\cal R} - \phi^{-7}{\cal A}+\phi^5\, {\cal C}\,,
	\end{equation}
where the shorthand ${\cal A}$ denotes the transverse-traceless components of the momentum tensor,
	\begin{equation}\label{ccM}
	{\cal A}=\frac{\pi^{ij}_{\,TT}\,\pi^{\,TT}_{ij}}{g},
	\end{equation}
and we have introduced the spatial constant ${\cal C}$, defined as
	\begin{equation}\label{cc9}
	{\cal C}:=\frac{1}{3\left(3\lambda-1\right)}\frac{\pi^2}{g}-\,2\,\Lambda\,.
	\end{equation}
As we will show, the sign of ${\cal C}$ determines the behaviour of eq.\ \eqref{cc7}. 

We assume that the scalar curvature ${\cal R}$ is a spatial constant for the reasons outlined in Sec.\ \ref{sec:gr} (see appendix \ref{app yam} for details). Recall that we have shown that using decomposition \eqref{cc3}, the only $\lambda$-dependent piece of the Legendre transformation lies on the trace of $\pi$. Hence, all the $\lambda$-dependent pieces of the modified Lichnerowicz-York equation are encoded in ${\cal C}$. Since we want to understand how the solutions of the constraints depend on $\lambda$, we proceed to discuss the properties of eq.\ \eqref{cc7} separately for ${\cal C}\geq 0$ and ${\cal C}<0$.

\subsubsection{Positive and vanishing ${\cal C}$}
The reason for discussing these two cases together is that establishing existence and uniqueness of solutions follows from the general relativistic analysis presented in Sec.\ \ref{sec:gr}, without the need for further considerations. If ${\cal C}=0$, eq.\ \eqref{cc7} reduces to
	\begin{equation}\label{cc10}
	8\,\nabla^2\,\phi=\phi \,{\cal R}-\phi^{-7}{\cal A}\,,
	\end{equation}
which is the Lichnerowicz equation \cite{AL}. It follows that as long as $g_{ij}$ belongs to the positive Yamabe class, there is a unique solution to eq.\ \eqref{cc10}. Note that for $\Lambda\neq 0$, ${\cal C}=0$ is only possible for a non-vanishing choice of $\frac{\pi}{\sqrt{g}}$. The constraint-solving data will therefore not resemble any set obtained from the original Lichnerowicz equation as a cosmological constant was not included then. We discuss this point further in Sec.\ \ref{Disc} when we compare the solutions of the $\lambda$-R model to those of general relativity. 

When, on the other hand, we have that ${\cal C}>0$, the existence and uniqueness of solutions can be deduced in a straightforward manner from the analysis of the original Lichnerowicz-York equation. We conclude that when ${\cal C}>0$ there almost always exist unique solutions to the modified Lichnerowicz-York equation and the Yamabe class of the initial data is not restricted\footnote{The set of restrictions associated with the ``almost always '' was discussed in Sec.\ \ref{sec:gr} and holds in the present case.}. 

To make this comparison between equations precise, consider a set of initial data $\big\{g_{ij},\pi^{ij}_{\,TT},\pi\big\}$ for the $\lambda$-R model and given values $\left\{\lambda^*,\Lambda^*\right\}$ for the constants $\lambda$ and $\Lambda$, such that ${\cal C}>0$. Denote the particular value of ${\cal C}$ for this configuration by ${\cal C}_0$, that is
	\begin{equation}\label{cc11}
	{\cal C}_0=\frac{1}{3\left(3\lambda^*-1\right)}\frac{\pi^2}{g}-2\Lambda^*>0\,.
	\end{equation}
Then, there exists a general relativity set of initial data $\big\{g_{ij},\pi^{ij}_{\,TT},\pi_0\big\}$, where $\pi^{ij}_{TT}$ is such that ${\cal A}$ is bounded, while $\pi_0$ satisfies the constant mean curvature condition and is such that
	\begin{equation}\label{cc12}
	\frac{1}{6}\left(\frac{\pi_0}{\sqrt{g}}\right)^2\!={\cal C}_0\,.
	\end{equation}
It follows that the same conformal factor $\phi$ that uniquely solves the Lichnerowicz-York equation with initial data $\big\{g_{ij},\pi^{ij}_{\,TT},\pi_0\big\}$ also uniquely solves the modified Lichnerowicz-York equation with initial data $\big\{g_{ij},\pi^{ij}_{\,TT},\pi\big\}$ and couplings $\left\{\lambda^*,\Lambda^*\right\}$. As we further elaborate after discussing the ${\cal C}<0$ case, the spacetimes evolved from the same conformal factor in both theories are in general different. For now, suffice it to say that for ${\cal C}>0$ there always exists a unique solution to eq.\ \eqref{cc7} regardless of the given (values of) initial data, while for ${\cal C}=0$ the initial spatial metric is restricted to belong to the positive Yamabe class.

\subsubsection{Negative ${\cal C}$}
In terms of the existence of solutions, the most interesting case happens when $\lambda$, $\Lambda$, and $\pi$ are such that ${\cal C}<0$. Recall that for the modified Lichnerowicz-York equation to have a solution, the polynomial $P(\phi)$ must have at least one zero. As we can see from the asymptotic behaviour of $P(\phi)$ when ${\cal C}<0$, this is not guaranteed in general. For the case ${\cal A}\neq 0$, we have
	\begin{equation}\label{AsymptP}
	\lim_{\phi\rightarrow \,0^+}P(\phi)=-\infty,\qquad \lim_{\phi\rightarrow +\infty}P(\phi)=-\infty,
	\end{equation}
while for the special case ${\cal A}=0$, we have
	\begin{equation}\label{AsymptPAz}
	\lim_{\phi\rightarrow \,0^+}P(\phi)=0,\qquad \lim_{\phi\rightarrow +\infty}P(\phi)=-\infty,
	\end{equation}
which also does not guarantee that there is a finite $\phi_0$ for which $P(\phi_0)$ vanishes.

\begin{figure}[h!]
\includegraphics[scale=0.75,center]{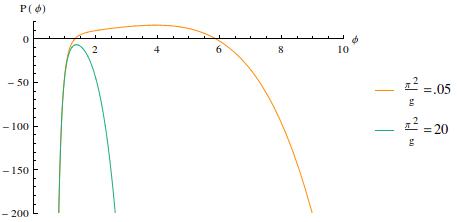}
\caption{Comparison of $P(\phi)$ for two different values of $\frac{\pi^2}{g}$, $\frac{\pi^2}{g}=0.5$ and $\frac{\pi^2}{g}=20$. The other parameters are kept fixed and are given by $\left({\cal R},{\cal A},\lambda,\Lambda\right)=\left(5,50,-1,0\right)$.}
\label{fig:f1}
\end{figure}
\begin{figure}[h!]\includegraphics[scale=0.75,center]{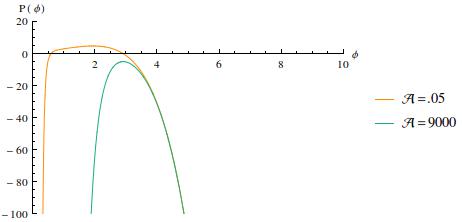}
\caption{Comparison of $P(\phi)$ for two different values of ${\cal A}$, ${\cal A}=0.5$ and ${\cal A}=9000$. The other parameters are kept fixed and are given by $\big({\cal R},\frac{\pi^2}{g},\lambda,\Lambda\big)=
\left(3,0.5,-1,0\right)$.}
\label{fig:f2}
\end{figure}

Note that the only possibly non-negative contribution to $P(\phi)$ comes from the term linear in $\phi$ and therefore the polynomial can only vanish for some $\phi_0>0$ if the initial metric belongs to the positive Yamabe class. Let us thus assume that ${\cal R}>0$ for the remainder of this subsection. This is a necessary condition to ensure the existence of a solution, but not a sufficient one. 

To see why this is so, let us briefly return to the vanishing ${\cal C}$ case. There, it was clear that if the metric $g_{ij}$ belonged to the positive Yamabe class, the large-$\phi$ behaviour of $P(\phi)$ was determined by the (positive) linear term, while as $\phi$ approached zero it was determined by the (negative) $\phi^{-7}$-term. The existence of a zero followed straightforwardly from the intermediate value theorem. When ${\cal C}<0$, the large-$\phi$ behaviour is no longer determined by the linear term but by the $\phi^{-7}$-term, which is now negative. As we can see from Figs.\ \ref{fig:f1} and \ref{fig:f2}, it is possible to change the number of zeros of $P(\phi)$ from two to zero by changing the values of $\pi^{ij}_{\,TT}$ and $\pi$. Hence, for a given initial choice of $\pi$ and $\pi^{ij}_{\,TT}$, ${\cal R}$ must be large enough for a bounded interval to exist in which $P(\phi)>0$.

From now on, let us assume that not only are $\lambda$, $\Lambda$, and $\pi$ such that ${\cal C}>0$, but ${\cal R}$ is also large enough to ensure that $P(\phi)$ has two zeros. As explained in subsection \ref{sec:YLe}, to apply the theorems guaranteeing the existence of a unique solution to the Lichnerowciz-York equation, there must exist a bounded interval $\left(\phi_-,\phi_+\right)$ such that $P\left(\phi_-\right)<0$ and $P\left(\phi_+\right)>0$ holds for all $x\in\Sigma_{t_0}$, with $\phi_-,\phi_+$ constants. It thus follows that we can only ensure the existence of a solution around the first zero of $P(\phi)$, which we henceforth denote by $\phi_1$.

To determine when does such an interval exist, suppose we are given some initial values of $\pi^{ij}_{\,TT}$ and $\pi$. Then, there must be a finite interval $I\ni \phi_+$ and for which
	\begin{equation}\label{cc13}
	\left\{\phi\in I\, | \,{\cal R}>{\cal A}\,\phi^{-8}\!-{\cal C}\phi^4\right\},
	\end{equation}
holds for every $x^i\in\Sigma_{t_0}$. Because ${\cal R}$ is a constant, for the set in eq.\ \eqref{cc13} to be non-empty, ${\cal A}$ must be bounded in $\Sigma_{t_0}$. Moreover, the minimal ${\cal R}$ for which eq.\ \eqref{cc13} is valid depends on the point $x^i\in\Sigma_{t_0}$. It is thus more convenient to write the inequality in terms of the maximum norm of the transverse-traceless initial data $A$,
	\begin{equation}
	A:=\max_{x^i\in\,\Sigma_{t_0}}{\cal A}\,.
	\end{equation}
We can then replace condition \eqref{cc13} by
	\begin{equation}\label{cc14}
	\left\{\phi\in I\, | \,{\cal R}>A\,\phi^{-8}\!-{\cal C}\phi^4\right\}.
	\end{equation}
This guarantees that ${\cal R}$ is large enough to ensure the existence of both zeros on all $\Sigma_{t_0}$, and that the position of the first zero is bounded from above. Further note that if $\pi^{ij}_{\,TT}=0$ for some $x^i\in\Sigma_{t_0}$, there is no $\phi_-<\phi_+$ such that $P\left(\phi_-\right)<0$, as required to ensure the existence of the solution. We must therefore impose that ${\cal A}$ is also bounded from below.

We have established three conditions that must be simultaneously satisfied for the modified Lichnerowicz-York equation to have at least one solution when ${\cal C}<0$:
	\begin{itemize}
	\item $g_{ij}$ belongs to the positive Yamabe class,
	\item ${\cal A}$ is bounded on $\Sigma$, and
	\item inequality \eqref{cc14} is satisfied.
	\end{itemize}
When all three of these conditions are satisfied, we are guaranteed that a solution exists around $\phi_1$, the smaller zero of $P(\phi)$. 

However, the polynomial has another zero, which we denote by $\phi_2>\phi_1$. Because its derivative with respect to $\phi$ at $\phi_2$ is negative, we cannot apply the theorem proved in \cite{YM2} around $\phi_2$. Nevertheless, the inapplicability of the theorem does not necessarily mean that no solution exists around $\phi_2$. We now show that in some limiting cases, a solution does exist.

\begin{figure}[h!]
\includegraphics[scale=0.75,center]{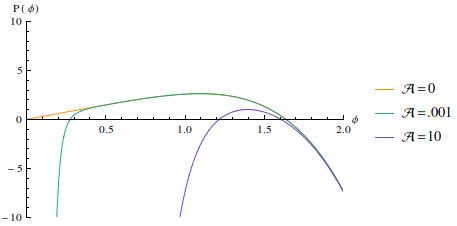}
\caption{Comparison of $P(\phi)$ for three different values of ${\cal A}$, ${\cal A}=0$, ${\cal A}=0.001$, and ${\cal A}=10$. The other parameters are kept fixed and are given by $\big({\cal R},\frac{\pi^2}{g},\lambda,\Lambda\big)=\left(3,5,-1,0\right)$.}
\label{fig:f3}
\end{figure}

Suppose that instead of bounded transverse-traceless initial data, we have $\pi^{ij}_{\,TT}=0$. In this case, $P(\phi)$ reduces to
	\begin{equation}\label{cc15}
	P(\phi)=\phi \,{\cal R} + \phi^5 \,{\cal C}\,,
	\end{equation}
and, since both ${\cal R}$ and ${\cal C}$ are constants, it follows that the constant $\phi_c$ given by
	\begin{equation}\label{cc16}
	\phi_c=\left(\frac{{\cal R}}{-{\cal C}}\right)^{1/4},
	\end{equation}
which exists as long as ${\cal R}>0$, is a solution to the generalised Lichnerowicz-York equation. Because we set the only non-constant term in $P(\phi)$ to zero, there is only one positive zero of $P(\phi)$ and it solves the equation, even though $P(\phi)$ is decreasing at this point. When we introduce a non-vanishing ${\cal A}$ which is such that condition \ref{cc14} is satisfied, $P(\phi)$ changes from having one zero at positive $\phi$ to two, and the solution whose existence we proved above exists around the new. Because we cannot apply the theorem proved in \cite{YM2}, let us consider ${\cal A}$ to be small enough that we can treat it as a perturbation around $\pi^{ij}_{\,TT}=0$. We then denote the infinitesimal $\phi^{-7}$-contribution to $P(\phi)$ by $\delta A>0$ and make the following replacements in eq.\ \eqref{cc7},
	\begin{subequations}\label{cc17}\begin{align}
	&{\cal A}\rightarrow \delta A,\\ &\phi\rightarrow \phi_c+\delta\phi,\label{cc17b}
	\end{align}\end{subequations}
reducing it to
	\begin{equation}\label{cc18}
	8\,\nabla^2\delta\phi=-4\,{\cal R}\,\delta \phi - \delta A\left(\frac{-{\cal C}}{{\cal R}}\right)^{7/4}.
	\end{equation}
This equation always has solutions as long as $\delta\phi<0$. The fact that the value of $\phi$ decreases when ${\cal A}\neq 0$ is explained by the fact that $\phi_2<\phi_c$ , as is illustrated in Fig.\ \ref{fig:f3}.

For the general situation in which the polynomial $P(\phi)$ has two zeros, $\phi_1$ and $\phi_2$, we have not been able to prove that two solutions always exist. What we managed to show is that for an initial $g_{ij}$ belonging to the positive Yamabe class, if ${\cal A}$ is bounded on $\Sigma_{t_0}$ and inequality \eqref{cc14} is satisfied, there exists a solution to the modified Lichnerowicz-York equation around $\phi_1$. When ${\cal A}$ vanishes, there is a unique constant solution at $\phi_c$ given in eq.\ \eqref{cc16}. Finally, when we introduce a small, non-vanishing ${\cal A}$, the first solution re-appears, in that case co-existing with a perturbed version of $\phi_c$, which is located around $\phi_2$. The behaviour of the polynomial around $\phi_2$ prevents us from using theorem 1 from \cite{YM2} to establish the existence of a second solution beyond the perturbative regime. Nevertheless, our perturbative results are suggestive that this is indeed the case.

\subsection{The lapse-fixing equation}\label{Lapse-Fix}
We now consider the quaternary constraint of the $\lambda$-R model, namely the lapse-fixing equation ${\cal M}\approx 0$,
	\begin{equation}
	{\cal M}\approx 0\Leftrightarrow {\cal R}-3\Lambda+\frac{a^2}{2\left(3\lambda-1\right)} N-  \frac{\dot{a}}{2}\approx  \nabla^2N,
	\end{equation}
where $a=\frac{\pi}{\sqrt{g}}$ is a function of time only. Because we are working with constraint solving variables, it is not guaranteed that ${\cal R}$ has a fixed sign. Hence, it is useful to re-write the equation using the Hamiltonian constraint,
	\begin{align}
	{\cal M}\approx 0\Leftrightarrow &\left[\left(\frac{a^2}{6\left(3\lambda-1\right)}+\frac{\pi^{ij}_{TT}\pi^{TT}_{ij}}{g}-\Lambda\right)N-\frac{\dot{a}}{2}\right)\approx \nabla^2 N\\
	\Leftrightarrow & \left(\left({\cal A} + \frac{{\cal C}}{2}\right)N-\frac{\dot{a}}{2}\right) \approx \nabla^2N,
	\end{align}
where in the last line we have adopted the shorthands ${\cal A}$ and ${\cal C}$ used in the previous sections. Notice that this ${\cal A}$ has, in principle, at different value than its initial data counterpart, while ${\cal C}$ is the same when written for initial and constraint-solving data since $\frac{\bar{\pi}^2}{g}=\frac{\pi^2}{g}$. 
By using Stokes' theorem on both sides of the equation, we see that the left-hand side must vanish at some $x$ in order for a solution to exist. This happens at 
	\begin{equation}
	N=\frac{\dot{a}}{2\left({\cal A}+\frac{{\cal C}}{2}\right)}.
	\end{equation}
Since $N>0$ by construction, this is only possible if the right-hand side is also positive, which is not guaranteed \emph{a priori}. When ${\cal C}\geq 0$, a solution to the lapse-fixing equation exists if $\dot{a}>0$. When ${\cal C}$ is negative, there are several options. The first is that while ${\cal C}<0$, the transverse-traceless data is such that ${\cal A}+\frac{{\cal C}}{2}>0$ and $\dot{a}>0$ is once again imposed. Alternatively, it can be that the transverse-traceless data is such that ${\cal A}+\frac{{\cal C}}{2}<0$ in the entire hypersurface, in which case $\dot{a}<0$ is imposed. When the sign of ${\cal A}+\frac{{\cal C}}{2}$ changes along the manifold, then there is a problem, since the sign of $\dot{a}$ and of $N$ must be constant in the manifold. Of course, if $\lambda=1$, this can occur in general relativity for a positive cosmological constant and certain choices of transverse-traceless data.

To conclude, the solution of the lapse-fixing equation will in general be $\lambda$-dependent, as not only ${\cal C}$ depends on $\lambda$, but so does the conformal factor obtained in the modified Lichnerowicz-York equation and therefore ${\cal A}$. However, the properties of the lapse-fixing equation in the $\lambda$-R model are the same as in general relativity.
\section{Discussion}\label{Disc}
As we have shown, there are three regimes that determine the restrictions on the initial data for the initial value formulation of the $\lambda$-R model. These regimes are defined by the sign of ${\cal C}$ at $\Sigma_{t_0}$, which depends on the value of three constants, namely, $\pi^2/g$, $\lambda$, and $\Lambda$. To compare the solutions of the $\lambda$-R model with those of general relativity, we begin by considering $\Lambda=0$. In this case, we have
	\begin{equation} {\cal C}=\frac{1}{3\left(3\lambda-1\right)}\frac{\pi^2}{g},\end{equation}
and it follows that sign of ${\cal C}$ is given by
	\begin{equation}
	\sign {\cal C}=\begin{cases} -1 & \mbox{if} \; \lambda<1/3,\\ 0 & \mbox{if} \; \pi=0,\\ +1 & \mbox{if} \; \lambda>1/3.\end{cases}
	\end{equation}
Notice that in this case, ${\cal C}$ only vanishes when $\pi=0$, regardless of the value of $\lambda$. As previously established, the generalised Lichnerowicz-York equation only admits solutions for ${\cal C}=0$ when the metric $g_{ij}$ belongs to the positive Yamabe class. Moreover, when $\pi=0$ $\lambda$ drops out of the Lichnerowicz-York equation and its solutions agree with general relativistic solutions written in the maximal slicing case. It is also true that when $\pi=0$ the equations of motion of the $\lambda$-R model coincide with those of general relativity in the maximal slicing gauge. 

\begin{figure}[h!]
\includegraphics[scale=0.75,center]{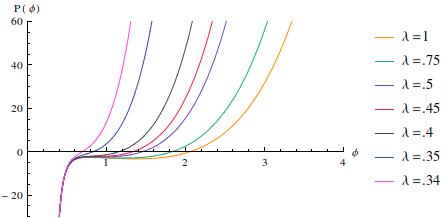}
\caption{Comparison of $P(\phi)$ as $\lambda$ decreases from $1$ to $\frac{1}{3}$. All other parameters are kept fixed and are given by $\big({\cal R},{\cal A},\frac{\pi^2}{g}\big)
=\left(-3,0.05,1\right)$.}
\label{fig:f4}
\end{figure}

When $\pi\neq 0$ and $\lambda>1/3$, we have ${\cal C}>0$ and therefore know that there is a unique solution for almost all possible choices of initial data, regardless of the Yamabe class of the initial metric. This allows us to address the behaviour of $\phi$ as $\lambda$ changes away from its general relativistic value of $1$. If we set the scalar curvature of $\Sigma_0$ to zero, ${\cal R}=0$, and impose $P(\phi)=0$, we obtain
	\begin{equation}\label{zero}
	-\phi^{-7}\!{\cal A}+\phi^5{\cal C}=0\;\Rightarrow\; {\cal A}={\cal C}\phi^{12}.
	\end{equation}
Further imposing ${\cal A}$ to be finite and non-vanishing means that the same must hold for the right-hand side of eq.\ \eqref{zero}. However, ${\cal C}$ goes to infinity as $\lambda$ approaches $\frac{1}{3}$ from above. Hence, to keep the product ${\cal C}\phi^{12}$ finite, $\phi$ must scale as
	\begin{equation}\label{cm3}
	\phi\propto {\cal C}^{\,^{-1}\!/\!_{12}}.
	\end{equation}
In other words, as ${\cal C}$ approaches infinity, the value $\phi_s$ for which $P(\phi_s)=0$ decreases according to eq.\ \eqref{cm3}, which is illustrated by Fig.\ \ref{fig:f4}. This behaviour is explained by the fact that the solution must be located in a finite neighbourhood of $\phi_s$. The argument can be extended to include a finite value for ${\cal R}$. Since the associated term in the modified Lichnerowicz-York equation is linear in $\phi$, the leading scaling behaviour is still dictated by eq.\ \eqref{cm3}.

It is also worth analysing the behaviour of the solutions when $\lambda\rightarrow \infty$. In this limit, ${\cal C}\rightarrow 0$, which leads to an increase on the position of the zero of $P(\phi)$. For given choices of ${\cal A}$ and $g_{ij}$, the maximum value of $\phi$ which solves the modified Lichnerowicz-York equation is obtained for ${\cal C}=0$, regardless of the value of $\pi$. In the exact ${\cal C}\rightarrow 0$ limit, one obtains a conformal factor which in general relativity would be associated with maximal slicing coordinates. This analysis shows that in the $\lambda$-R model, when $\lambda\rightarrow\infty$, its physics are effectively described by a maximal slicing configuration despite the fact that both its initial and constraint-solving data include a non-vanishing constant mean curvature $\pi$. To illustrate that this is indeed true for all times, we turn to the time evolution equations for $\bar{g}_{ij}$ and $\bar{\pi}^{ij}_{\,TT}$,
	\begin{subequations}\label{cm4}\begin{align}
	&\dot{\bar{g}}_{ij}=\frac{2N}{\sqrt{\bar{g}}}\left(\bar{\pi}^{\,TT}_{ij}-\frac{\bar{g}_{ij}\,\bar{\pi}}
	{3\left(3\lambda-1\right)}\right),\\
	&\dot{\bar{\pi}}^{ij}_{\,TT}=\frac{N}{\sqrt{\bar{g}}}\left(\frac{2}{3\left(3\lambda-1\right)}\,\bar{\pi}^{ij}_{\,TT}
	\bar{\pi}-2\,\bar{g}_{kl}\,\bar{\pi}^{ik}_{\,TT}\,\bar{\pi}^{jl}_{\,TT}\right)-\sqrt{\bar{g}}\,N\left(\bar{{\cal R}}^{ij}-\frac{1}{3}\bar{g}^{ij}\bar{{\cal R}}\right)
	\lp&\qquad\,\,\,\,-\sqrt{\bar{g}}\left(\bar{g}^{ik}\bar{g}^{jl}-\frac{1}{3}\bar{g}^{ij}\bar{g}^{kl}\right)\bar{\nabla}_k\bar{\nabla}_l N,\label{cm4b}
	\end{align}\end{subequations}
where $\bar{\nabla}_i$ denotes the covariant derivative with respect to $\bar{g}_{ij}$. Despite the arbitrary value of $\bar{\pi}$, in the $\lambda\rightarrow\infty$ limit, all such contributions vanish from the equations of motion, which therefore match those of general relativity in the maximal slicing gauge.

From the last two paragraphs, it is clear that providing the same initial data to both the original and modified Lichnerowicz-York equation, the resulting conformal factor differs and is $\lambda$-dependent. As we discuss below, that does not prove that the theories are inequivalent as the initial data is not physical and it is possible to match the constraint-solving data of both theories by relating the initial data in a $\lambda$-dependent way.
\begin{figure}[h!]
\includegraphics[scale=0.45,center]{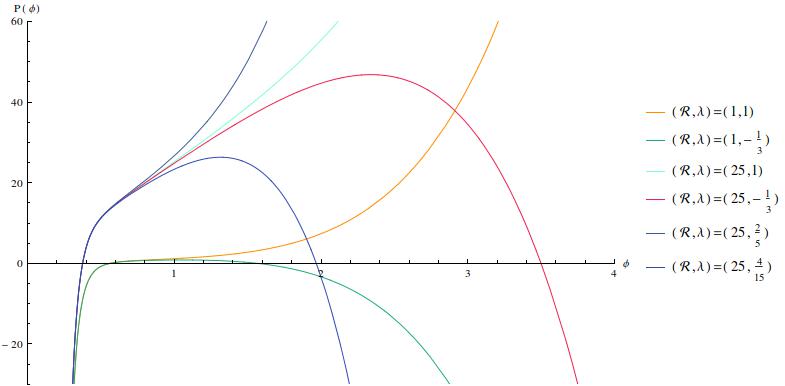}
\caption{Comparison of $P(\phi)$ for pairs of $\lambda$ yielding the same $\left|{\cal C}\right|$ and different values of ${\cal R}$ . Every other parameter was kept fixed.}
\label{fig:f5}
\end{figure}
We now consider the ${\cal C}<0$ ($\lambda <1/3$) case, which for $\Lambda=0$ does not occur in general relativity. As we established in Sec.\ \ref{IVLR}, the existence of solutions requires that for given $\pi$ and ${\cal A}$, the initial base metric $g_{ij}$ belongs to the positive Yamabe class and is such that ${\cal R}$ is large enough to satisfy relation \eqref{cc14}. Consider a set of initial data for which there is at least one solution\footnote{Recall that in Sec.\ \ref{IVLR} above, we discussed some cases for which there are two solutions.} and denote the respective value of ${\cal C}<0$ by ${\cal C}*$,
	\begin{equation}
	{\cal C}*=\frac{1}{3\left(3\lambda-1\right)}\frac{\pi^2}{g}<0.
	\end{equation}
Select the same initial data and a constant $\lambda'$ such that ${\cal C}$ is positive and has the value ${\cal C}=\left|{\cal C}*\right|$,
\begin{figure}[h!]
\includegraphics[scale=0.375,center]{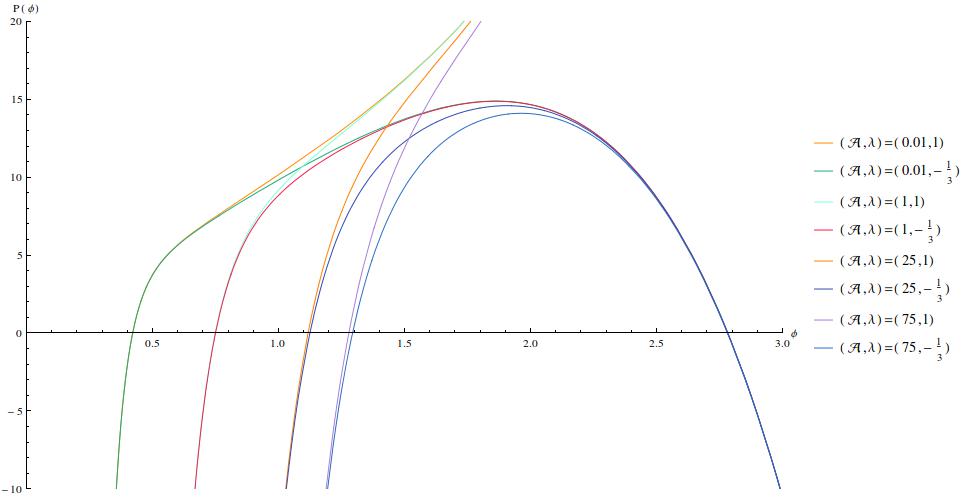}
\caption{Comparison of $P(\phi)$ for pairs of $\lambda$ yielding the same $\left|{\cal C}\right|$ and different values of ${\cal R}$ . Every other parameter was kept fixed.}
\label{fig:f6}
\end{figure}
	\begin{equation}
	\left|{\cal C}*\right|=\frac{1}{3\left(3\lambda'-1\right)}\frac{\pi^2}{g}>0,\qquad \Rightarrow \qquad \lambda'=\frac{2}{3}-\lambda.
	\end{equation}
For ${\cal C}={\cal C}*$, the first zero of the polynomial $P(\phi)$ occurs when the linear term $\phi{\cal R}$ compensates the $\phi^{-7}$-term and before the $\phi^5$-term becomes the dominant one. The same is true for the only zero of $P(\phi)$ when ${\cal C}=\left|{\cal C}*\right|$, since the only difference between the polynomials is the sign of the $\phi^5$ term, whose magnitude is the same. It follows that the solution to the modified Lichnerowicz-York equation around $\phi_1$ for ${\cal C}={\cal C}*$ will be close to the solution for the ${\cal C}=\left|{\cal C}*\right|$. 

It follows that the constraint-solving data is approximately the same for small $\phi_1$. However, notice that the equations of motion for $\bar{g}_{ij}$ and $\bar{\pi}^{ij}_{TT}$ will have a sign difference in the $\bar{\pi}$-terms. Nevertheless, because $P(\phi)$ only depends on $\pi^2$, there will be an approximate matching in the time evolution of the constraint solving data obtained from $\left(g_{ij},\pi^{ij}_{TT},\pi,{\cal C}*\right)$ and from $\left(g_{ij},\pi^{ij}_{TT},-\pi,\left|{\cal C}*\right|\right)$. As Figs.\ \ref{fig:f5} and \ref{fig:f6} show, this approximate matching becomes less and less valid as ${\cal C}$ increases. 

We now consider a non-vanishing cosmological constant $\Lambda$. Notice that, in the presence of a cosmological constant, ${\cal C}$ only vanishes when $\pi\neq 0$. More concretely, when $\lambda>1/3$ ($\lambda<1/3$), $\Lambda$ must be positive (negative) and the initial value of $\pi$ is given by
	\begin{equation}\label{VanCLam}
	\frac{\pi^2}{g}=6\left(3\lambda-1\right)\Lambda.
	\end{equation}
We therefore find yet another case in which ${\cal C}$ vanishes but $\pi$ does not satisfy the maximal slicing condition. Unlike the $\lambda\rightarrow\infty$ limit, this is a possible regime in general relativity as eq.\ \eqref{VanCLam} above is valid for positive $\Lambda$ when $\lambda=1$. However, while in general relativity one can substitute the $\frac{\bar{\pi}}{\sqrt{\bar{g}}}$-terms in the equations of motion \eqref{cm4} by $\sqrt{\Lambda}$, these terms remain explicitly $\lambda$-dependent in the $\lambda$-R model, as the $\left(\left(3\lambda-1\right)\Lambda\right)^{1/2}$ does not cancel the $\left(3\lambda-1\right)^{-1}$ factors in the equations of motion. Another way of understanding this case is to notice that when $\pi$ satisfies eq.\ \eqref{VanCLam}, then eq.\ \eqref{cm4b} describes the evolution of a transverse-traceless momentum density in the constant mean curvature gauge of general relativity with an effective trace term $\bar{\pi}_{eff}$ given by
	\begin{equation}\label{cm5}
	\bar{\pi}_{eff}=\frac{2}{3\lambda-1}\,\bar{\pi}=\frac{2\,\phi^6}{3\lambda-1}\,\pi\,.
	\end{equation}
The previously discussed effects of $\lambda\neq 1$ in the solutions of the Lichnerowicz-York equation are not altered when $\Lambda\neq 0$, with the exception of the $\lambda\rightarrow\infty$ limit, in which ${\cal C}=-2\Lambda$ instead of ${\cal C}=0$. Finally, the range of $\lambda$ for which ${\cal C}$ is of a certain sign depends on $\Lambda$. For $\Lambda>0$, we have
	\begin{equation}
	\sign {\cal C}=\begin{cases} -1 & \mbox{if} \; \lambda<\frac{1}{3},\;\;\mbox{or}\;\;\lambda >\frac{6\Lambda+\frac{\pi^2}{g}}{18\Lambda}\\ 0 & \mbox{if} \; \lambda=\frac{6\Lambda+\frac{\pi^2}{g}}{18\Lambda},\\ +1 & \mbox{if} \; \frac{1}{3}<\lambda<\frac{6\Lambda+\frac{\pi^2}{g}}{18\Lambda},\end{cases}
	\end{equation}
while for $\Lambda<0$,
	\begin{equation}
	\sign {\cal C}=\begin{cases} -1 & \mbox{if} \; \frac{6\Lambda+\frac{\pi^2}{g}}{18\Lambda}<\lambda<\frac{1}{3},\\ 0 & \mbox{if} \; \lambda=\frac{6\Lambda+\frac{\pi^2}{g}}{18\Lambda},\\ 
	+1 & \mbox{if} \; \lambda<\frac{6\Lambda+\frac{\pi^2}{g}}{18\Lambda}\;\;\mbox{or}\;\; \lambda>1/3.\end{cases}
	\end{equation}
Below, we establish the conditions under which the constraint-solving data of the $\lambda$-R model and of general relativity are the same, and compare the time evolution of that set of data according to the equations of motion of each theory.

As we have seen, whenever we are able to match the constraint-solving data of the $\lambda$-R model with that of general relativity, the evolution equations are manifestly different and therefore the theories with those choices of initial data are not equivalent. However, we can attempt to match the evolution data, which means matching the constraint-solving data $\bar{g}_{ij}$ and $\bar{\pi}^{ij}_{\,TT}$ in both theories, while relating the trace terms via
	\begin{equation}\label{67}
	\bar{\pi}_{\lambda R}=\frac{3\lambda-1}{2}\,\bar{\pi}_{GR}\,.
	\end{equation}
Since we are matching constraint-solving data, we can take the barred $\lambda$-R Hamiltonian constraint and write it as a function of $\phi_{GR}$ and the remaining general relativistic initial data, using eq.\ \eqref{67} for the trace term. We thus obtain the following version of the modified Lichnerowicz-York equation,
	\begin{equation}\label{68}
	8\nabla^2\phi_{GR}=\phi_{GR}\, {\cal R}-\phi^{-7}_{GR}\,{\cal A}
	+\phi^{5}_{GR}\left(\frac{3\lambda-1}{12}\frac{\pi^2}{g}-2\Lambda\right).
	\end{equation}
Using the fact that $\phi_{GR}$ solves the usual Lichnerowicz-York equation, we obtain
	\begin{equation}\label{69}
	\phi^5_{GR}\,\frac{\pi^2}{g}\frac{\lambda-1}{4}=0\,,
	\end{equation}
which is only true if either $\lambda=1$ or $\pi=0$, the two cases already known to yield equivalence between the theories. Similarly, we can also allow for a constant additive shift between the two cosmological constants, because there is no reason to assume that both models should be written with the same value of the cosmological constant. Setting $\Lambda_{\lambda R}=\Lambda_{GR}+\Lambda'$ effectively turns equation \eqref{68} into
	\begin{equation}\label{70}
	\phi^5_{GR}\left(\frac{\pi^2}{g}\frac{\lambda-1}{4}-2\Lambda'\right)=0\,.
	\end{equation}
Since $\frac{\pi}{\sqrt{g}}$ is a spatial constant, there always exists a $\Lambda'$ such that eq.\ \eqref{70} is valid on $\Sigma_0$. Imposing eq.\ \eqref{70} does not spoil the matching of time evolution, because the cosmological constant drops out from the equations. Moreover, although $\frac{\pi}{\sqrt{g}}$ is in general a function of time, eq.\ \eqref{70} only refers to the initial data and therefore to its value at that particular point in time. In terms of comparing the initial value formulations of both models, this would imply including either $\lambda$ or $\Lambda'$ in the initial data. For general values of these couplings, there is no way to match both theories unless one fine-tunes the values of these parameters as we have just illustrated.

In summary, we have studied the initial value formulation of the $\lambda$-R model by applying the conformal method developed by Lichnerowicz, York and \'{O} Murchadha, which is particularly suited as its underlying condition $\nabla_i\pi=0$ is a constraint of the $\lambda$-R model. Analogous to what happens in general relativity, the Hamiltonian constraint becomes an equation for the conformal factor of the metric, which we referred to as the modified Lichnerowicz-York equation \eqref{cc7}. This equation differs from its $\lambda=1$ counterpart only in the $\phi^5$-term, which we denoted by ${\cal C}$. In the absence of a cosmological constant, the range of ${\cal C}$ therefore differs from its general relativistic counterpart. More importantly, for given values of $\frac{\pi}{\sqrt{g}}$ and $\Lambda$, the sign of ${\cal C}$ is $\lambda$-dependent.

For vanishing ${\cal C}$, the solutions to the modified equation are the same as those of the traditional one for initial data obeying the maximal slicing condition and base metric in the positive Yamabe class. We further argued that unless $\pi=0$ and $\Lambda=0$, the time evolution of the model does not match that of general relativity in the constant mean curvature gauge, since the equations of motion for $g_{ij}$ and $\pi^{ij}_{\,TT}$ depend on $\pi$ and $\lambda$ in a manifestly different way.

For positive ${\cal C}$, the existence and uniqueness of solutions follows straightforwardly from the general relativistic case. We argued that in the limit $\lambda\rightarrow 1/3$ (and therefore ${\cal C} \rightarrow \infty$), the conformal factor scales as $\phi\propto {\cal C}^{\,^{-1}\!/\!_{12}}$. When $\lambda\rightarrow\infty$ and ${\cal C}\rightarrow 0$, general relativity is recovered since $\lambda$ drops out of the equation. We have also explained how it is possible to scale the initial data in order to have the same constraint-solving data both in the $\lambda$-R model and in general relativity. This makes explicit that, unless $\pi=0$, the constraint surfaces match only at the initial hypersurface, because the time evolution of both theories is manifestly different. In addition, we have shown that the only way to obtain matching constraint-solving data whose time evolution is the same is for either $\lambda=1$ or $\pi=0$.

Finally, we studied the case of negative ${\cal C}$. This regime can occur when $\lambda=1$, if $\Lambda$ is large enough compared to the choice of $\frac{\pi}{\sqrt{g}}$. Similar to the case of vanishing ${\cal C}$, only metrics belonging to the positive Yamabe class can yield solutions. Even then, the allowed choices of base metric depend on the initial value of the momentum tensor, since the spatial curvature ${\cal R}$ must be large enough for solutions to exist. We have shown that for a bounded choice of transverse-traceless initial data there always exists a solution. When $\pi^{ij}_{\,TT}=0$ everywhere on $\Sigma$, there is a constant solution to the equation regardless of the value of ${\cal R}$ (as long as it is admissible). Moreover, for a very small but non-vanishing $\pi^{ij}_{\,TT}$, perturbative arguments show that a small (negative) perturbation around the constant solution remains a solution, coexisting with the one mentioned previously.

Comparing general relativity and the $\lambda$-R model for ${\cal C}<0$ is more subtle than for ${\cal C}>0$, although the conclusions are similar, namely those pertaining to the non-equivalence between solutions. The only way to have ${\cal C}<0$ in general relativity is when $\Lambda$ is sufficiently large. In this case the conditions that ${\cal R}$ must be sufficiently large and that $\pi^{ij}_{\,TT}$ must be bounded still apply, and solutions can be found. Naturally, one can fine-tune $\pi$ and $\Lambda$ to find the same value of ${\cal C}$, regardless of the value of $\lambda$. However, for the same reason that no equivalence was obtained when ${\cal C}>0$ unless $\pi=0$, no equivalence is found here.

\vspace{.5cm}

\noindent {\bf Acknowledgements.} 
LP acknowledges financial support from Funda\c{c}\~{a}o para a Ci\^{e}ncia e Tecnologia, Portugal through 
grant no. SFRH//BD/76630/2011. I would also like to express my gratitude to professor Niall \'{O} Murchadha for hosting my in Cork in 2015 and helping me considerably in the undertaking of the research underlying this paper. Moreover, I would also like to thank professor R. Loll and S. Gryb for helpful discussions during various stages of this work.

\vspace{1cm}

\begin{appendices}
\section{Yamabe classification of Riemannian manifolds}\label{app yam}
In this appendix, we briefly review some results originally obtained by Japanese mathematician Hidehiko Yamabe \cite{Y1,Y2} regarding conformal properties of Riemannian manifolds. He showed that for a Riemannian manifold $\Sigma$ which is either compact or asymptotically flat, with dimension $d\geq 3$, and equipped with a metric $g_{ij}$, there always exists a conformal transformation taking $g_{ij}$ to $\tilde{g}_{ij}$ such that the Ricci scalar associated with $\tilde{g}_{ij}$ is constant. Moreover, there is a conformally invariant constant, since dubbed the ``Yamabe constant'' $Y$, and defined as
	\begin{equation}\label{c10}
	Y=\inf\frac{\int d^3x\sqrt{g} \left(R\theta^2+8\left(\nabla \theta\right)^2 \right)}{\left(
	\int d^3x\sqrt{g} \,\theta^6\right)^{1/3}},
	\end{equation}
whose sign defines a conformal equivalence class of metrics. When the minimising function $\theta$ is used as a conformal factor, $\Sigma$ is mapped to a manifold of constant curvature denoted by $R_0$. Once the transformation is done, the same invariant $Y$ can be computed with $\theta=1$ as a minimizing function, yielding
	\begin{equation}\label{c11}
	Y=R_0V_0^{2/3},
	\end{equation}
where $V_0=\int d^3x\sqrt{g}$ is the volume of $\Sigma$. The sign of $Y$ tells us that the manifold can be conformally mapped to another with constant curvature of the same sign. This splits all metrics into three Yamabe classes, defined by having positive, negative and vanishing Yamabe constants.

One could then change the notation and denote the arbitrary initial data by $(\tilde{g}_{ij},\tilde{K}_{ij}^{TT})$. Before moving towards the constraint solving data, one first computes the Yamabe constant and performs a conformal transformation by $\theta(x)$ into a set $\left(g_{ij},K^{TT}_{ij}\right)$ whose scalar curvature is now constant and from then proceeds as described above until obtaining the Lichnerowicz equation, this time around with the linear term being a constant. However, to avoid adding yet another set of variables, we will remain with our nomemclature of initial data for the set $\left(g_{ij},K^{TT}_{ij}\right)$ and assume a constant scalar curvature, knowing that in terms of uniqueness and existence of solutions, it represents a whole class of initial data.

\end{appendices}

\end{document}